\documentclass[aps,prl,reprint,superscriptaddress]{revtex4-1}
\usepackage{graphicx,epsfig}
\usepackage{amssymb}
\usepackage{amsmath}
\usepackage{bm}
\usepackage{textcomp}
\usepackage{color}
\usepackage{float}
\usepackage{comment}
\usepackage{mathtools}

\usepackage{soul}

\begin{document}
\setcounter{page}{1}

\title[]{Moving crystal phases of a quantum Wigner solid in an ultra-high-quality 2D electron system}
\author{P. T. \surname{Madathil}}
\affiliation{Department of Electrical and Computer Engineering, Princeton University, Princeton, New Jersey 08544, USA}
\author{K. A. \surname{Villegas Rosales}}
\affiliation{Department of Electrical and Computer Engineering, Princeton University, Princeton, New Jersey 08544, USA}
\author{Y. J. \surname{Chung}}
\affiliation{Department of Electrical and Computer Engineering, Princeton University, Princeton, New Jersey 08544, USA}
\author{K. W. \surname{West}}
\affiliation{Department of Electrical and Computer Engineering, Princeton University, Princeton, New Jersey 08544, USA}
\author{K. W. \surname{Baldwin}}
\affiliation{Department of Electrical and Computer Engineering, Princeton University, Princeton, New Jersey 08544, USA}
\author{L. N. \surname{Pfeiffer}}
\affiliation{Department of Electrical and Computer Engineering, Princeton University, Princeton, New Jersey 08544, USA}
\author{L. W. \surname{Engel}}
\affiliation{National High Magnetic Field Laboratory, Florida State University, Tallahassee, Florida 32310, USA}
\author{M. \surname{Shayegan}}
\affiliation{Department of Electrical and Computer Engineering, Princeton University, Princeton, New Jersey 08544, USA}

\date{\today}

\begin{abstract}

In low-disorder, two-dimensional electron systems (2DESs), the fractional quantum Hall states at very small Landau level fillings ($\nu$) terminate in a Wigner solid (WS) phase, where electrons arrange themselves in a periodic array. The WS is typically pinned by the residual disorder sites and manifests an insulating behavior, with non-linear current-voltage (\textit{I-V}) and noise characteristics. We report here, measurements on an ultra-low-disorder, dilute 2DES, confined to a GaAs quantum well. In the $\nu < 1/5$ range, superimposed on a highly-insulating longitudinal resistance, the 2DES exhibits a developing fractional quantum Hall state at $\nu=1/7$, attesting to its exceptional high quality, and dominance of electron-electron interaction in the low filling regime. In the nearby insulating phases, we observe remarkable non-linear \textit{I-V} and noise characteristics as a function of increasing current, with current thresholds delineating three distinct phases of the WS: a pinned phase (P1) with very small noise, a second phase (P2) in which $dV/dI$ fluctuates between positive and negative values and is accompanied by very high noise, and a third phase (P3) where $dV/dI$ is nearly constant and small, and noise is about an order of magnitude lower than in P2. In the depinned (P2 and P3) phases, the noise spectrum also reveals well-defined peaks at frequencies that vary linearly with the applied current, suggestive of washboard frequencies. We discuss the data in light of a recent theory that proposes different dynamic phases for a driven WS.
\end{abstract}

\maketitle  

Two-dimensional electron systems (2DESs) in a sufficiently strong, perpendicular magnetic field ($B$) at low temperatures and very small Landau level filling factors ($\nu < 1/5$), form a many-body, ordered array of electrons, known as the quantum Wigner solid (WS) \cite{wigner1934interaction,lam1984liquid,levesque1984crystallization,yi1998laughlin}. The electrons, having their kinetic energy quenched significantly, succumb to the repulsive Coulomb interaction, and form a triangular lattice, maximizing the distance between each pair. In the presence of disorder, the WS breaks into domains and gets pinned to the local disorder sites which significantly alter the bulk transport properties \cite{cha1994orientational,zhu1994sliding,fisher1979defects,ruzin1992pinning,shayegan1997perspectives,chitra1998dynamical}. The longitudinal resistance ($R_{xx}$) of a pinned WS exhibits an insulating, typically activated behavior, $R_{xx} \propto e^{E_A/2k_BT}$ \cite{jiang1991magnetotransport}, where $E_A$ is commonly associated with the WS defect formation energy \cite{narevich2001hamiltonian,archer2014quantum,chung2022correlated,madathilunpublished}. The pinned WS displays microwave or radio-frequency resonances that can be understood as collective excitations caused by the oscillations of the WS domains within the pinning potential \cite{andrei1988observation,chen2006melting}. It can also show non-linear current-voltage (\textit{I-V}) and noise characteristics, revealing the complex dynamics of a moving WS \cite{shayegan1997perspectives,willett1989current,goldman1990evidence,williams1991conduction,li1991low,jiang1991magnetotransport,li1995rf,li1996inductive,csathy2007astability}.

While the presence of disorder is a crucial ingredient in the transport properties of a pinned WS, less disordered 2DESs exhibit additional remarkable structure in transport measurements. Recent magnetoresistance measurements in ultra-high-quality 2DESs \cite{chung2021ultra} indeed show clear signatures of fragile fractional quantum Hall states (FQHSs) at $\nu = 1/7$ and other fillings embedded deep within the insulating phase \cite{chung2022correlated}, clearly signaling the presence of interaction-driven phenomena at very small fillings. One would expect the quality to be reflected in the WS phase as well, and yield novel transport signatures. Our Letter reports striking non-linear \textit{I-V} and noise characteristics of the driven WS in this extremely high-quality regime.

The ultra-high-quality 2DES in our study is realized by confining electrons to an 89-nm-wide GaAs quantum well. The sample was grown using molecular beam epitaxy, following an extensive campaign to improve the crystal purity via an optimization of the growth environment and sample structure \cite{chung2021ultra,chung2022understanding}. The quantum well has flanking Al$_x$Ga$_{1-x}$As barriers with stepped values of $x$, and Si dopants inside a doping-well structure \cite{chung2021ultra,chung2022correlated, chung2020working,chung2022understanding}. The very dilute 2DES has an areal density of $\simeq 2.83 \times 10^{10}$ cm$^{-2}$ and a record-high mobility of $\simeq 15\times 10^6$ cm$^2$/Vs at such low density. We performed electrical transport measurements on a $4 \times 4$ mm$^2$ van der Pauw geometry, with alloyed In:Sn contacts at its corners and edge midpoints. The sample was cooled  in a dilution refrigerator. For details of our measurement protocols, see Supplemental Material \cite{SupplementalMaterial}.

\begin{figure}[h]
\centering
\includegraphics[width=0.48\textwidth,height=0.89\linewidth]{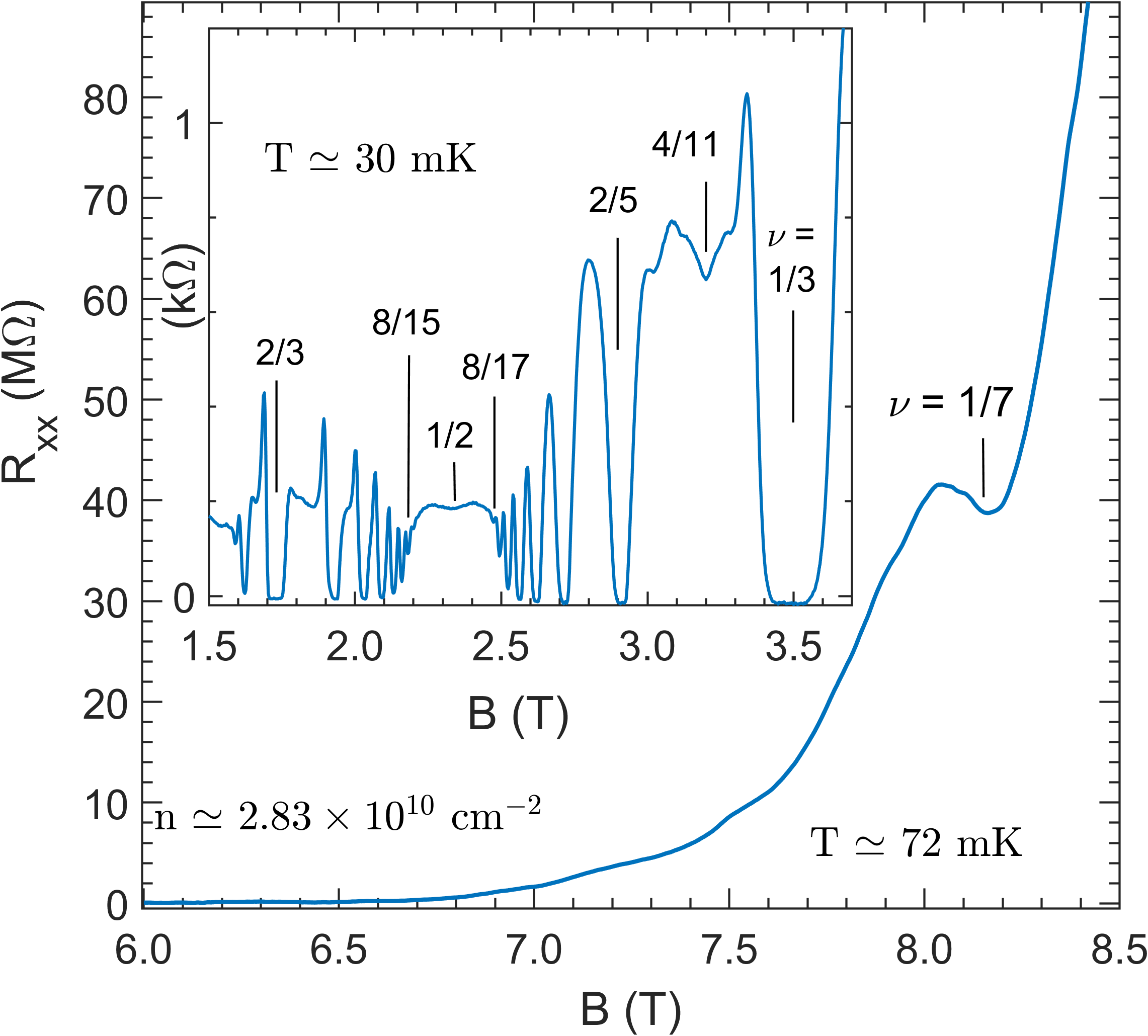}
\centering
  \caption{\label{Ip} 
Longitudinal resistance ($R_{xx}$) vs. magnetic field ($B$) for a dilute 2DES confined to an 89-nm-wide GaAs quantum well at $T\simeq$ 72 mK in the small filling range. The position of an emerging FQHS at $\nu = 1/7$ is marked. Inset: $R_{xx}$ at $T\simeq$ 30 mK in the higher filling range.
  }
  \label{fig:Ip}
\end{figure}

Figure 1 presents $R_{xx}$ vs. $B$ for our 2DES. The sample's exceptionally high quality is evinced from the well-developed FQHSs up to $\nu = 8/17$ and $\nu = 8/15$ on the flanks of $\nu = 1/2$ (Fig. 1 inset), as well as an emerging FQHS between $\nu = 1/3$ and 2/5, at $\nu = 4/11$, which can be understood as the FQHS of \textit{interacting} composite fermions \cite{wojs2000quasiparticle,wojs2004fractional,mukherjee2014enigmatic}. The $\nu = 4/11$ FQHS is extremely fragile and has been reported before only in the highest quality 2DESs and at much higher densities \cite{pan2015fractional,samkharadze2015observation,chung2021ultra,chung2022correlated}. Its presence in our dilute 2DES is a testament to its unprecedented quality. The data in the high-field range (Fig. 1 main) show that the resistance is approximately four orders of magnitude larger than in the low-field trace, indicating a highly insulating behavior when the filling factor is small ($\nu < 1/5$). Nevertheless, we can see a clear $R_{xx}$ minimum at $\nu = 1/7$, strongly corroborating the high quality of this dilute 2DES.

 Figure 2(a) displays the non-linear transport traces at four different temperatures below the melting temperature of the WS which we deduce to be $T_m \simeq 120$ mK \cite{SupplementalMaterial,footnote.melting,deng2019probing}. We measure the differential resistance ($dV/dI$) as a function of the driving \textit{dc} current ($I$). A small sinusoidal ($ac$) excitation current of 0.05 nA is superimposed on $I$, and the ratio of the measured differential voltage to the small signal excitation gives the value of \textit{dV/dI}. The frequency for the $ac$ current is kept small (0.1 Hz) to maximize the in-phase (resistive) component. The data are obtained at $B$ = 8.40 T ($\nu = 0.139$), past the $\nu = $1/7 FQHS. 

 At the lowest $T$ [80 mK, top trace in Fig. 2(a)], $dV/dI$ vs. $I$ exhibits a strong non-linear behavior which can be broken into three distinct regimes, separated by two threshold currents. At very small $I$, $dV/dI$ drops fairly sharply, by a factor of about 5 within $\simeq$1 nA, and then shows a very abrupt drop at a first threshold current ($I_{th1}$) to negative values. We associate $I_{th1}$ with the depinning of the pinned WS. For $I > I_{th1}$, $dV/dI$ becomes non-monotonic and fluctuates around zero. Beyond a second threshold current ($I_{th2}$), the amplitude of the fluctuations significantly diminishes, and $dV/dI$ attains a small, relatively constant value, with little dependence on current. We associate the three different regimes separated by $I_{th1}$ and $I_{th2}$ with three different dynamic phases of the WS; we refer to these as P1, P2, and P3, as a function of increasing $I$, and show later in the manuscript that they each possess a distinct noise spectrum. 

 \begin{figure}[t]
\centering
\includegraphics[width=0.48\textwidth,height=1.30\linewidth]{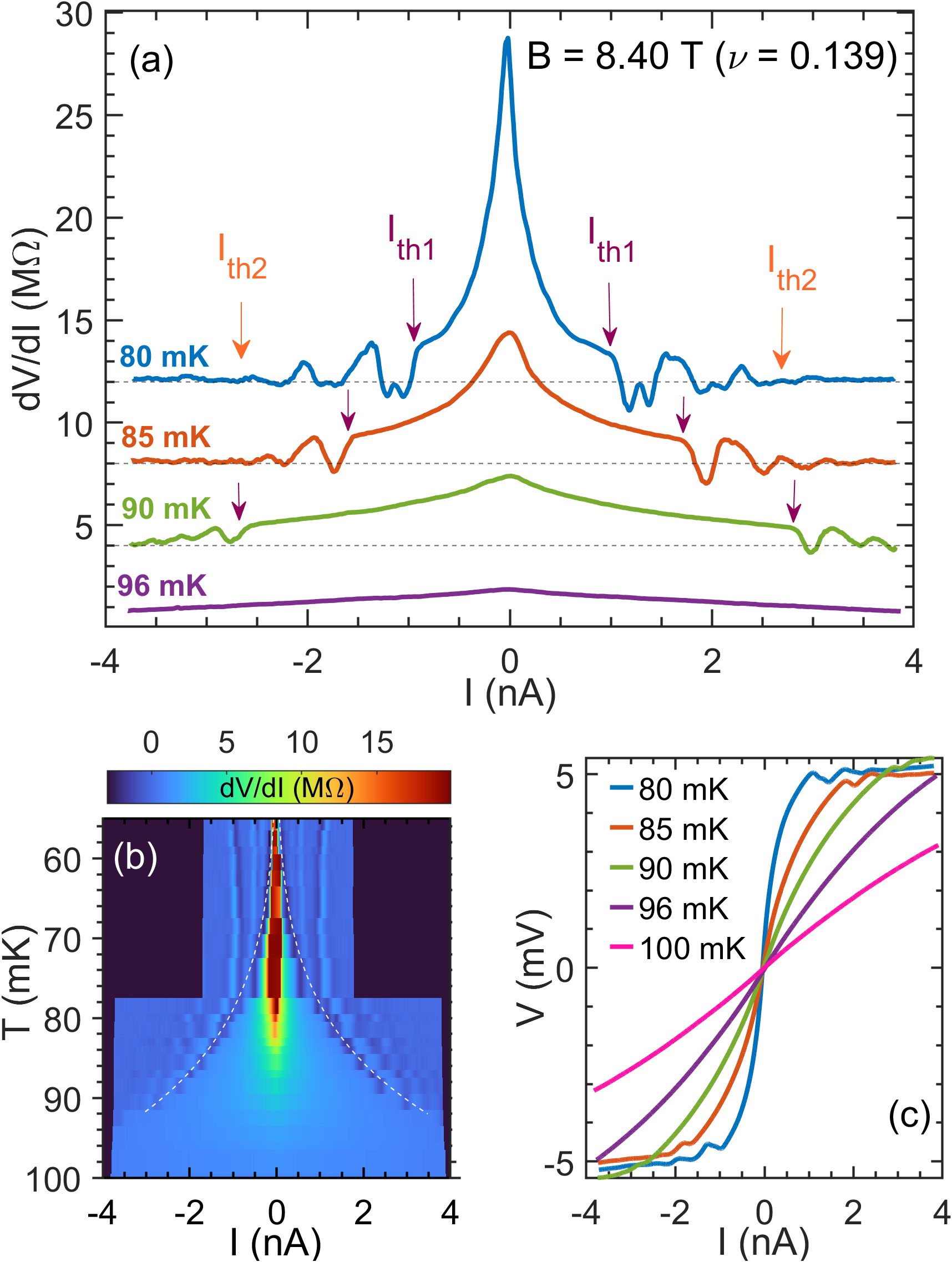}
\centering
  \caption{\label{Ip} 
(a) $dV/dI$ vs. $I$ at $B$ = 8.40 T ($\nu = 0.139$). The different traces, corresponding to different temperatures $T \simeq 80, 85, 90,$ and 96 mK, are offset vertically by 12, 8, 4, and 0 M$\Omega$, respectively. (b)  Color-scale plot summarizing $dV/dI$ vs. $I$ curves for all temperatures. The dashed white line follows the first minimum corresponding to the negative differential resistance seen just beyond $I_{th1}$. (c) The integrated \textit{I-V} characteristics at different temperatures.
  }
  \label{fig:Ip}
\end{figure}

\begin{figure*}[t]
\centering
\includegraphics[width=1\textwidth,height=0.5\linewidth]{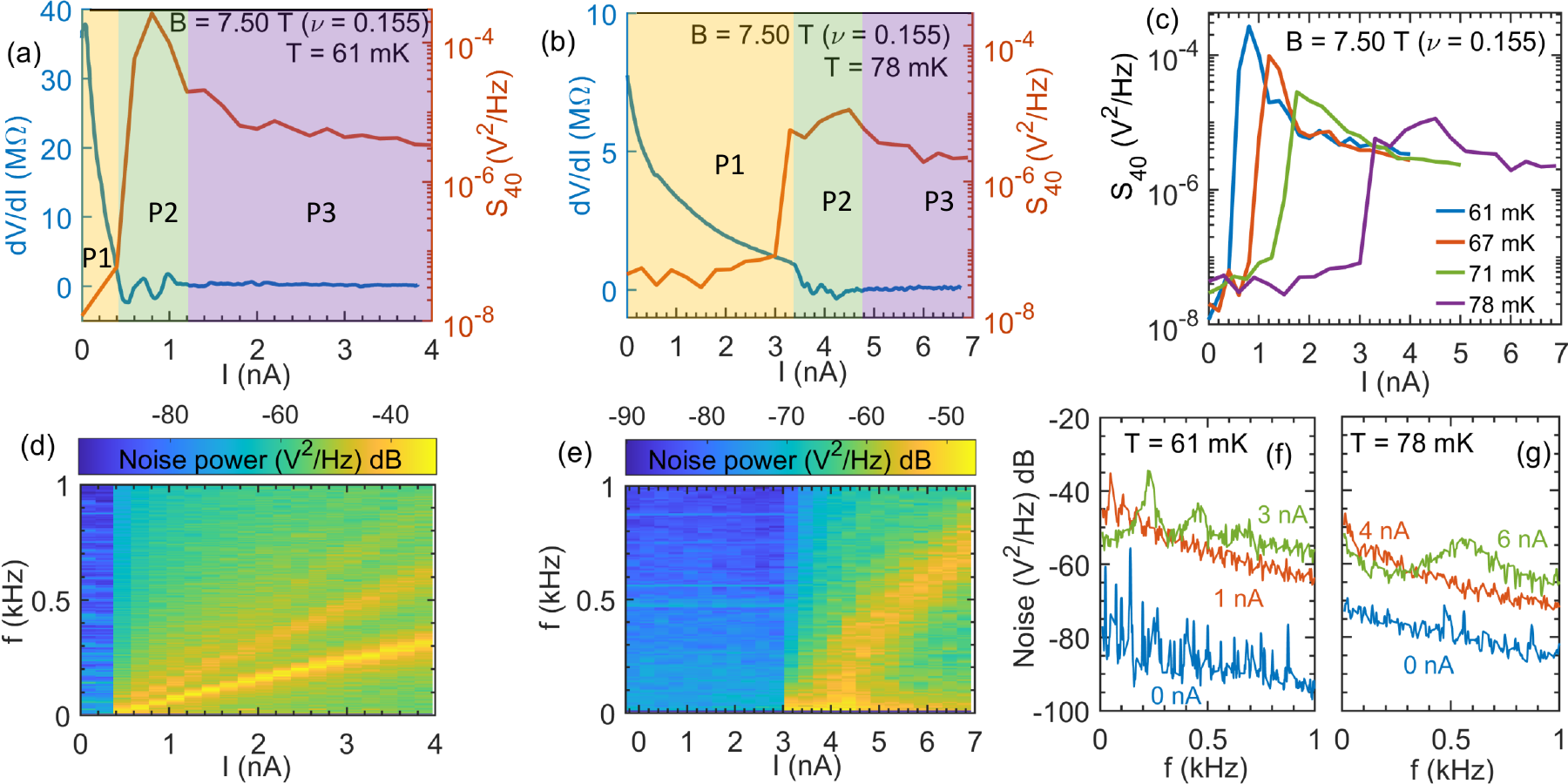}
\centering
  \caption{\label{Ip} 
 Noise data at $B$ = 7.50 T ($\nu$ = 0.155). (a,b) $dV/dI$ (blue trace) and noise power obtained by averaging over a frequency window of $\pm$10 Hz centered at 40 Hz, defined as $S_{40}$ (red trace), vs. $I$ at $T$ $\simeq$ 61 and $\simeq$ 78 mK, respectively. (c) $S_{40}$ vs. $I$ at different temperatures. The scale for $S_{40}$ is logarithmic and spans over four decades. (d,e) Color-scale plots summarizing the noise power as a function of $f$ and $I$ at $T$ $\simeq$ 61 and $\simeq$ 78 mK, respectively. (f,g) Noise power vs. $f$ for different values of $I$ as indicated.
}
  \label{fig:Ip}
\end{figure*}

 The traces shown in Fig. 2(a) reveal that the main peak becomes broader and $I_{th1}$ shifts towards higher values as $T$ increases. In addition, the local minima past $I_{th1}$, corresponding to negative $dV/dI$, become less negative and the amplitude of the fluctuations diminishes until it fully disappears at $T$ = 96 mK. Figure 2(b) is a color-scale plot that summarizes $dV/dI$ vs. $I$ curves at all temperatures. Figure 2(c) shows the \textit{I-V} characteristics obtained by integrating the $dV/dI$ vs. $I$ curves. At the lowest temperature, \textit{I-V} is step-like, with a large voltage developed for a small amount of current. This sharp jump can be understood as the voltage, or electric field, required to overcome the pinning potential and cause motion of the WS. Beyond $I_{th1}$, the voltage fluctuates about the threshold voltage value, and increases very gradually with $I$, indicating that once depinned, a negligible amount of extra electric field is required to sustain a current. Increasing temperature makes the step-like transitions more smooth, and the fluctuations become weaker, eventually vanishing at the highest temperatures \cite{vth.footnote}.

To further investigate the driven phases of the WS, we measured the noise spectrum in response to a driving $dc$ current, concomitantly with $dV/dI$. Figure 3 summarizes the noise data at $B$ = 7.50 T ($\nu$ = 0.155). In Fig. 3(a), we show the non-linear $dV/dI$ (blue trace) together with the noise power averaged between 30 and 50 Hz, defined as $S_{40}$ (red trace) vs. $I$ at $T\simeq$ 61 mK. Qualitatively similar to Fig. 2(a) data, the $dV/dI$ vs. $I$ data in Fig. 3(a) exhibit two current thresholds that delineate three phases. (Note that the data in Figs. 2(a) and 3(a) were taken at different fillings and temperatures.) There is a remarkable correlation between the behaviors of $S_{40}$ and $dV/dI$ with $I$ in Fig. 3(a) in the three phases. At low drives ($I < I_{th1} \simeq 0.4$ nA), the noise is low ($S_{40} < 10^{-7}$ V$^2/$Hz). We identify this pinned, low-noise phase as P1. For currents between $I_{th1}$ and $I_{th2}$ ($0.4 \lesssim I \lesssim 1.2$ nA), where $dV/dI$ fluctuates, $S_{40}$ rises dramatically, by about four orders of magnitude, and reaches a peak value of $\simeq$ $10^{-4}$ V$^2/$Hz. This is the second phase, P2. Past P2, in phase P3, $dV/dI$ is very low and constant, noise decreases and, for large driving currents ($I \gtrsim 2$ nA), it becomes roughly constant and settles at a value of $\simeq$ $3\times10^{-6}$ V$^2/$Hz.   

In Fig. 3(b) we present non-linear $dV/dI$ and $S_{40}$ data at a higher temperature of $T\simeq$ 78 mK. The $dV/dI$ trace indicates larger threshold currents, with suppressed fluctuating $dV/dI$ features in the P2 phase. The latter is also reflected in $S_{40}$ data, where the peak in $S_{40}$ in phase P2 is strongly suppressed compared to Fig. 3(a) data. Data shown in Fig. 3(c) at intermediate temperatures display the evolution of $S_{40}$ with temperature.

Besides $S_{40}$, we also measured the noise power spectrum between 5 Hz and 1 kHz as a function of frequency ($f$) at different driving currents. Figure 3(f) contains such data at $T\simeq$ 61 mK, for $I$ = 0, 1, and 3 nA. A summary of the full noise response vs. $f$ and $I$ at $T\simeq$ 61 mK is presented as a color-scale plot in Fig. 3(d). The noise spectrum reveals a distinct peak, accompanied by additional harmonics, that shift to higher frequencies with increasing $I$ in the P2 and P3 phases. The peak frequencies are linear in $I$, consistent with a signal that is of the washboard type \cite{gruner1988dynamics}. However, the frequency values we observe are $\sim$$10^{3}$ times smaller than the expected washboard frequencies estimated from the expression $f = J/nea_0$ where $J$ is the current density, $n$ is the electron density, $e$ is the electron charge and $a_0$ is the WS lattice constant \cite{li1995rf,gruner1988dynamics}. Figure 3(g) contains the noise response vs. $f$ at a higher temperature of $T\simeq$ 78 mK, for $I$ = 0, 4, and 6 nA. A summarized color-scale plot is presented in Fig. 3(e). At this higher temperature, only a broad peak is observed for $I$ = 6 nA. In the Supplemental Material \cite{SupplementalMaterial}, we report the temperature dependence of the slope of the $f$ vs. $I$ plots [see Figs. 3(d) and (e)], and of the power associated with the fundamental frequency of the narrow-band noise spectra.

\begin{figure}[t]
\centering
\includegraphics[width=0.4\textwidth,height=0.59\linewidth]{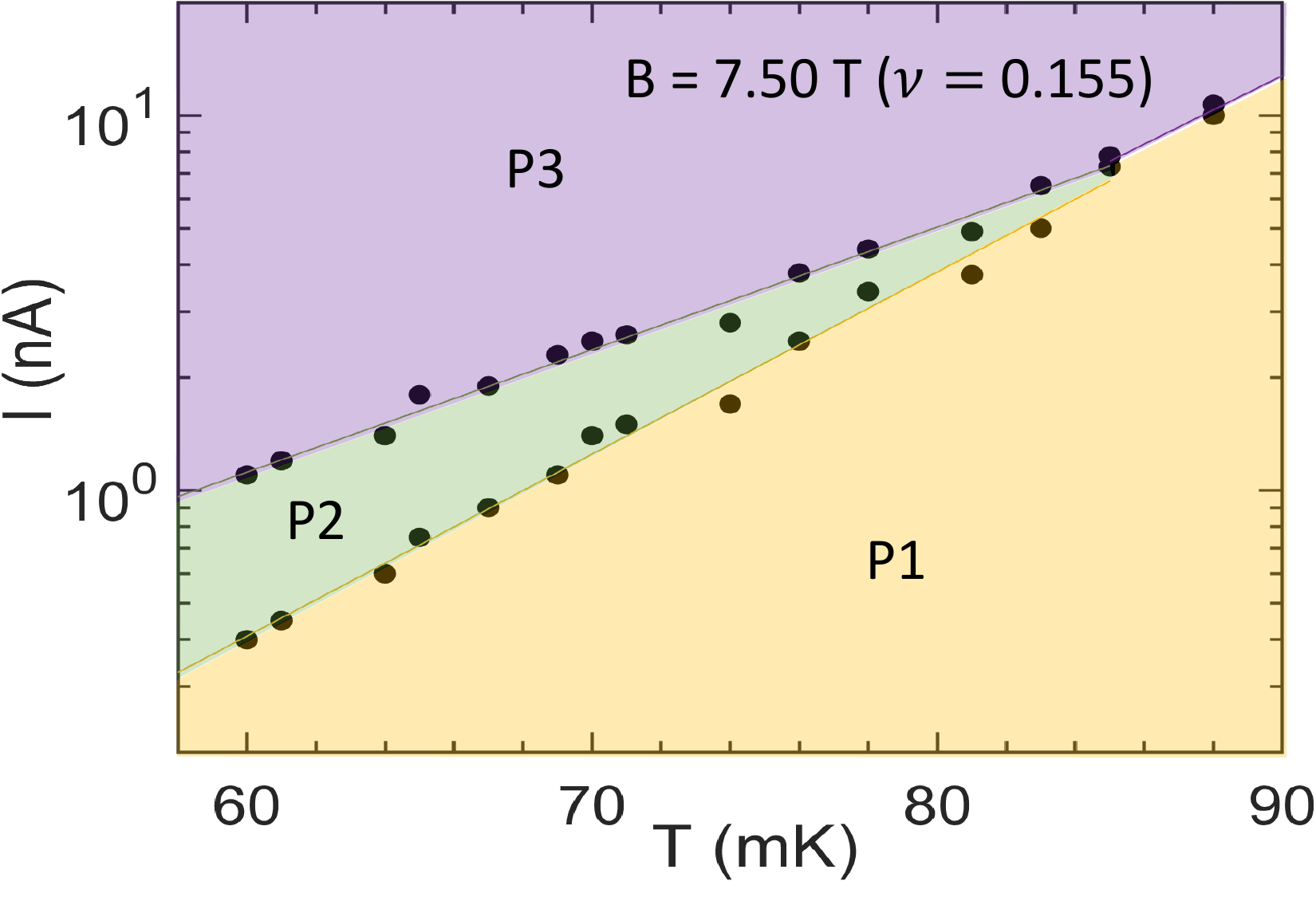}
\centering
  \caption{\label{Ip} 
Phase diagram for the WS as a function of $T$ and $I$ at $B$ = 7.50 T ($\nu$ = 0.155).
  }
  \label{fig:Ip}
\end{figure}

Our experimental observations can be summarized in a phase diagram for the WS as presented in Fig. 4. The data are taken at $B$ = 7.50 T ($\nu$ = 0.155), and $I$ is presented in a log-scale to visualize the width of the P2 region clearly. The black data points are obtained from the two thresholds, $I_{th1}$ and $I_{th2}$, extracted from $dV/dI$ vs. $I$ data. The extent of the P2 phase in $I$ is widest at the lowest temperatures. It narrows as $T$ increases and completely disappears above $T \simeq$ 85 mK (see Supplemental Material \cite{SupplementalMaterial} for $dV/dI$ and $S_{40}$ data at $T$ $\simeq85$ mK).  

Our observations of noise and the $dV/dI$ data follow closely the numerical calculations of Reichhardt and  Reichhardt \cite{reichhardt2001moving,reichhardt2022nonlinear,reichhardt2023noise}, which suggest the presence of a pinned phase, followed by a moving WS glass phase, and then a moving smectic phase, with increasing drive. At low temperatures, below the melting temperature of the WS, the disordered WS exhibits channel flow just above the first threshold. In this phase, the conduction is filamentary, and can show jumps in $dV/dI$ which can be negative \cite{reichhardt2001moving,reichhardt2022nonlinear,reichhardt2023noise}. Our $dV/dI$ data also show fluctuations, with negative $dV/dI$ just above depinning which disappears at $T \simeq$ 85 mK. Note that $T_m \simeq$ 120 mK for the WS in our sample \cite{SupplementalMaterial}, and thus the P2 phase is lost before the WS melts. Our experimental observation of fluctuating and negative $dV/dI$ in phase P2 is highly suggestive of the filamentary phase reported in Refs. \cite{reichhardt2022nonlinear,reichhardt2023noise}. 

Reichhardt and  Reichhardt   \cite{reichhardt2022nonlinear,reichhardt2023noise} also report on noise power (which we associate with $S_{40}$ in our work) as a function of the drive. Their calculations show suppressed noise in the pinned phase, high noise in the moving glass phase, and low noise in the moving smectic phase as the drive increases. Our $S_{40}$ data also reveal a similar qualitative picture, with $S_{40}$ being very low in phase P1, high in phase P2, and suppressed again in phase P3, with increasing current.

Despite the above qualitative agreements, there are some notable departures. While theory \cite{reichhardt2022nonlinear,reichhardt2023noise} predicts the emergence of the washboard signal only in the P3 (moving smectic) phase,  we observe the noise peaks in phases P2 and P3. Also, although the linear dependence of the peak frequency on $I$ that we observe [Fig. 3(d)] is consistent with the washboard frequency, the three orders of magnitude discrepancy between the experimental and estimated frequencies is puzzling and leaves room for speculation. One possibility is that the origin of the noise signal is the motion of WS domains and its interaction with the disorder potential. If we assume that the domains have a reasonably uniform size, and since the domain size is much larger than the WS lattice constant, the resulting peak frequencies can be much smaller than the washboard frequency \cite{monceau2012electronic}. Based on our observed $E_{th}\simeq$ 1 V/m, and the expression $L_0^2\simeq$ $0.02e/4\pi\epsilon\epsilon_0E_{th}$ \cite{fukuyama1978dynamics}, we obtain a rough estimate of $nL_0^2\simeq$ 600 electrons per domain \cite{footnote.domain,lili2023dynamic,matthew2023characterization,footnote.gaps}. It is worth mentioning that narrow-band-noise peaks have been reported in the reentrant integer quantum Hall states, and there too, the observed frequencies are much smaller than the expected washboard value \cite{cooper2003observation,sun2022dynamic}. In the Supplemental Material \cite{SupplementalMaterial}, we also report voltage ($V$) vs. time ($t$) traces of the noise measured with an oscilloscope. The real-time waveforms are saw-tooth like, suggesting a stick-slip motion of the domains as they encounter disorder sites \cite{gruner1994density,sekine2004sliding}.

There have been previous experiments reporting the non-linear \textit{I-V} and noise characteristics of the high-field ($\nu<$1/5) WSs \cite{csathy2007astability,li1991low,williams1991conduction,goldman1990evidence,jiang1991magnetotransport,willett1989current}. While the data presented in these studies have distinct dissimilarities, a common conclusion has been the existence of two phases, a pinned phase at low drives, followed by a sliding phase at high drives. In our ultra-high-quality samples, we report an additional phase where $dV/dI$ is negative and fluctuating, and is accompanied by a large, low-frequency noise [Fig. 3(c)]. This is different from the data in previous work; so is the presence of a current-dependent peak in the noise spectrum in the depinned phases [Figs. 3(d) and (e)] \cite{Footnote.Csathy}.

We also note that a rich collection of experiments on the negative $dV/dI$ exists in different platforms, suggesting a variety of possible mechanisms. In Refs. \cite{zou2021dynamical,rees2016stick,glasson2001observation}, e.g., transport measurements of classical WSs formed on the surface of superfluid helium at zero magnetic field are reported. Reference \cite{brussarski2018transport} reports two-threshold \textit{I-V} characteristics past the metal-insulator transition in Si-MOSFETs (metal-oxide-semiconductor field-effect transistors). While the phenomena reported in Refs. \cite{zou2021dynamical,rees2016stick,glasson2001observation,brussarski2018transport} happen in the insulating phases, a more surprising observation is the similarity of our \textit{I-V} features to those reported in Ref. \cite{bykov2007zero} in a GaAs 2DES, but in vastly different parameter ranges: $\sim$ 30 times larger density, $\sim$ 100 times higher temperatures, $\sim$ 10,000 times larger currents, and $\sim$ 10 times lower magnetic fields ($B\sim$ 0.8 T, $\nu \sim$ 50)! The \textit{I-V} characteristics were attributed to a redistribution of electrons in energy space, induced by the current \cite{bykov2007zero}. These observations highlight the complexities and yet certain qualitative similarities of the electrical transport properties in remarkably different platforms and circumstances. Finally, it has been theoretically proposed that composite fermion correlations present in the FQHSs persist in the WS phase, and that composite fermion WSs are more stable than electron WSs at very low fillings \cite{archer2013competing,archer2014quantum}. Our data presented here should prove useful in testing future theories dealing with the transport characteristics of composite fermion WSs.

We acknowledge support by the National Science Foundation (NSF) Grant No. DMR 2104771 for measurements. For sample characterization, we acknowledge support by the U.S. Department of Energy Basic Energy Office of Science, Basic Energy Sciences (Grant No. DEFG02-00-ER45841) and, for sample synthesis, NSF Grants No. ECCS 1906253 and the Gordon and Betty Moore Foundation’s EPiQS Initiative (Grant No. GBMF9615 to L.N.P.). We thank S. A. Lyon for loaning us the HP-3562A spectrum analyzer, and D. A. Huse and J. K. Jain for illuminating discussions.




\begin{thebibliography}{99}
\bibitem{vth.footnote} It is worth noting that, while the current threshold $I_{th1}$ increases with increasing temperature, the voltage threshold remains approximately fixed at $\simeq$ 5 mV [see Fig. 2(c)].
\end{thebibliography}
\end{document}


\setcounter{page}{1}

\title[]{Supplemental Material for ``Moving crystal phases of a quantum Wigner solid in an ultra-high-quality 2D electron system"}
\author{P. T. \surname{Madathil}}
\affiliation{Department of Electrical and Computer Engineering, Princeton University, Princeton, New Jersey 08544, USA}
\author{K. A. \surname{Villegas Rosales}}
\affiliation{Department of Electrical and Computer Engineering, Princeton University, Princeton, New Jersey 08544, USA}
\author{Y. J. \surname{Chung}}
\affiliation{Department of Electrical and Computer Engineering, Princeton University, Princeton, New Jersey 08544, USA}
\author{K. W. \surname{West}}
\affiliation{Department of Electrical and Computer Engineering, Princeton University, Princeton, New Jersey 08544, USA}
\author{K. W. \surname{Baldwin}}
\affiliation{Department of Electrical and Computer Engineering, Princeton University, Princeton, New Jersey 08544, USA}
\author{L. N. \surname{Pfeiffer}}
\affiliation{Department of Electrical and Computer Engineering, Princeton University, Princeton, New Jersey 08544, USA}
\author{L. W. \surname{Engel}}
\affiliation{National High Magnetic Field Laboratory, Florida State University, Tallahassee, Florida 32310, USA}
\author{M. \surname{Shayegan}}
\affiliation{Department of Electrical and Computer Engineering, Princeton University, Princeton, New Jersey 08544, USA}

\date{\today}

\maketitle 





\section{I.  E\lowercase{xperimental setup for} \boldmath{$dV/dI$}, \lowercase{noise, and Voltage (}$V$) \lowercase{vs. time (\textit{t}) measurements}}
\setlength{\parindent}{5mm}
\vspace{3mm}

\begin{figure}[h]
\centering
\includegraphics[width=0.5\textwidth]{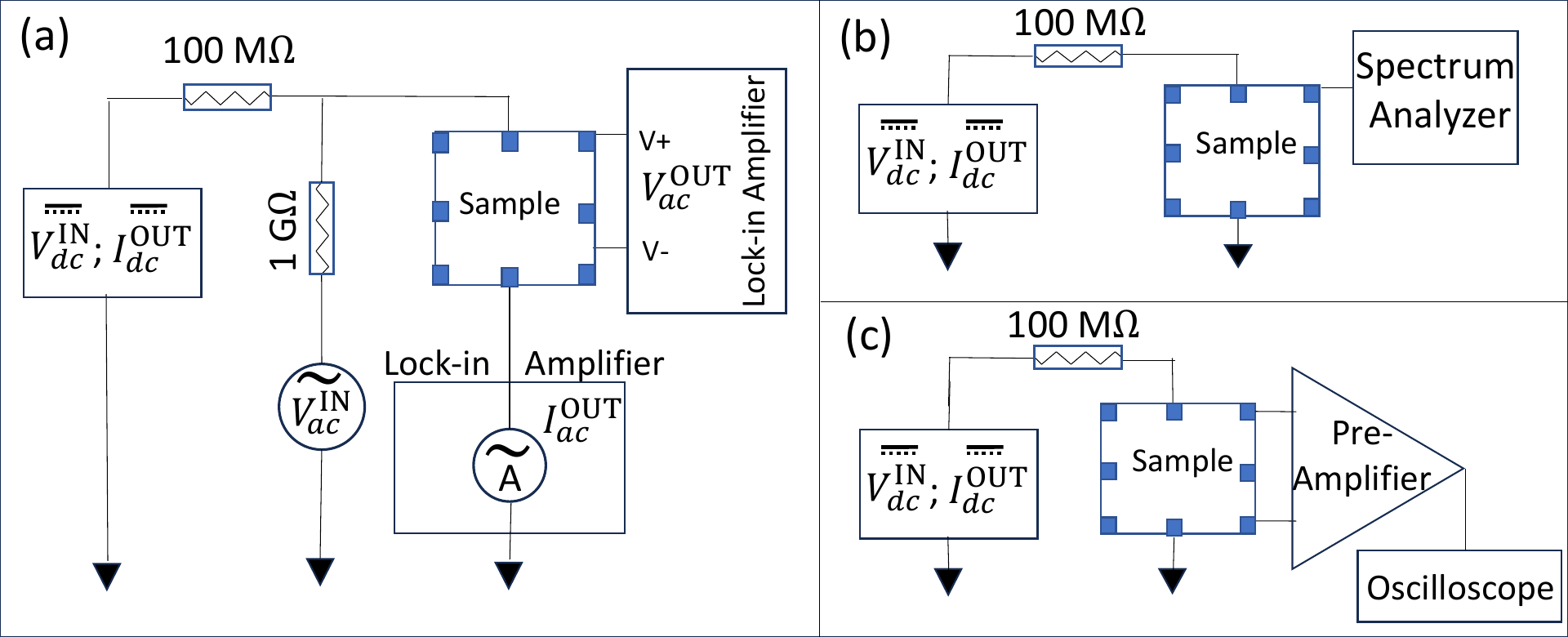}
\centering
  \caption{\label{Ip} 
  Measurement setup for: (a) $dV/dI$ vs. $I$, (b) noise spectrum, and (c) $V$ vs. $t$. 
  }
  \label{fig:Ip}
\end{figure}

Figure S1(a) shows the schematic for the $dV/dI$ vs. $I$ measurements. A voltage supply (Keithley 2410) is used to source a $dc$ voltage and measure the $dc$ current. The circuit is driven in the constant-current mode by connecting a 100 M$\Omega$ resistor in series to the sample. Additionally, an $ac$ voltage source with a 1 G$\Omega$ resistor is used to source the $ac$ excitation current for the differential resistance measurements. The ratio of $V_{ac}^{OUT}$ and $I_{ac}^{OUT}$ measured using lock-in amplifiers gives the value of $dV/dI$. The noise measurement setup is presented in Fig. S1(b). The protocol for the $dc$ drive remains the same. We use a spectrum analyzer (HP 3562A) to measure the noise power spectral density. Figure S1(c) shows the scheme for $V$ vs. $t$ measurements. A pre-amplifier is used to boost the voltage signal from the sample in the $ac$ coupling mode which is then monitored with an oscilloscope.

\section{II.  M\lowercase{elting temperature of the} W\lowercase{igner solid}}

\begin{figure}[h]
\centering
\includegraphics[width=0.5\textwidth]{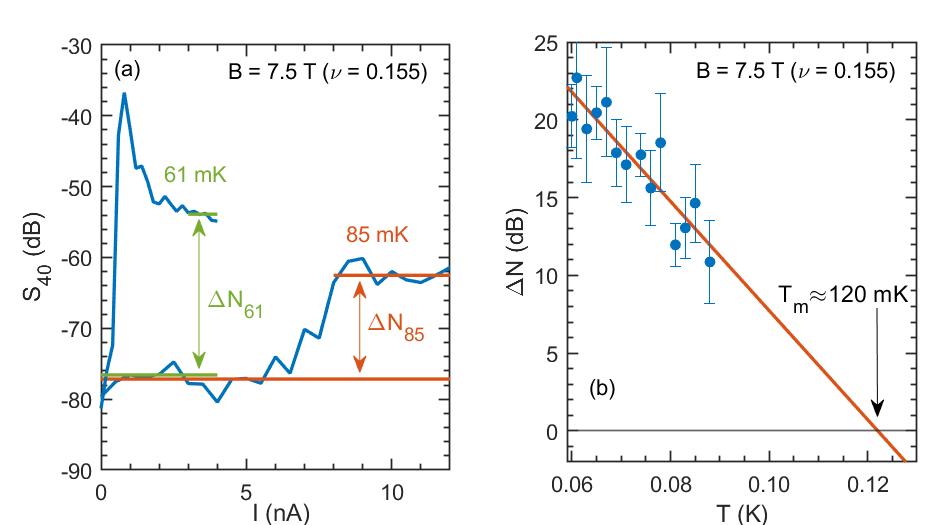}
\centering
  \caption{\label{Ip} 
 (a) $S_{40}$ data at $T$ $\simeq$ 61 and 85 mK at $B$ = 7.50 T ($\nu = 0.155)$ as a function of the driving current ($I$). The difference between the averaged noise power in the P1 and P3 phases is marked as $\Delta N_{61}$ and $\Delta N_{85}$ for the two different temperatures. (b) $\Delta N$ vs. $T$. From the x-intercept of the linear fit (red line), we estimate a melting temperature, $T_m \simeq$ 120 mK.
  }
  \label{fig:Ip}
\end{figure}

In our measurements, we can deduce the melting temperature ($T_m$) of the Wigner solid (WS) from the temperature dependence of the $S_{40}$ noise in the high-current regime (P3 phase). Figure S2(a) contains $S_{40}$ data at $T$ $\simeq$ 61 and 85 mK. The average $S_{40}$ power in the P1 phase (near $I$ = 0), and in the P3 phase, where $S_{40}$ saturates, are marked by the green lines for $T=61$ mK, and red lines for $T=85$ mK. We define the power difference between the P3 and P1 phases as $\Delta$N. Figure S2(b) shows the $\Delta$N values for various temperatures. The error bars are computed from the standard deviation of the $S_{40}$ data points involved in obtaining the average noise powers in the P1 and P3 phases. We observe that, despite the error bars, $\Delta$N falls off approximately linearly with $T$, and extrapolates to zero at a temperature which we associate with $T_m$. By fitting a straight line to the data points in Fig. S2(b), we estimate $T_m$ $\simeq$ 120 mK from the x-intercept of the fitted line.

\section{III.  \boldmath{$dV/dI$} \lowercase{and} $S_{40}$ \lowercase{vs.} \boldmath{$I$} \lowercase{at} $B$ = 7.50 T ($\nu$ = 0.155) \lowercase{and} \textit{T} $\simeq$ 85 \lowercase{m}K}
\setlength{\parindent}{0mm}

\begin{figure}[h]
\centering
\includegraphics[width=0.5\textwidth]{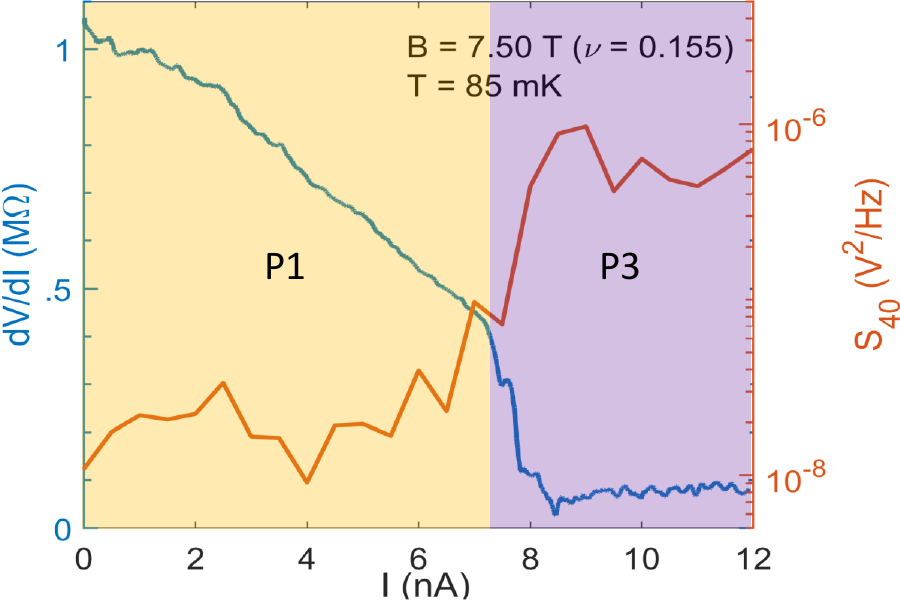}
\centering
  \caption{\label{Ip} 
 $dV/dI$ (blue trace) and noise power, defined as $S_{40}$ (red trace) vs. $I$ at $T$ $\simeq$ 85 mK. 
  }
  \label{fig:Ip}
\end{figure}
\vspace{2mm}

Figure S3 shows the $dV/dI$ and noise power, defined as $S_{40}$ in the main text, vs. $I$ at a high temperature of $T$ $\simeq$ 85 mK. The system makes a transition from the pinned phase to the constant, linear phase directly, without exhibiting any fluctuating or negative $dV/dI$ features. $S_{40}$ data also support this, revealing a single transition from a low-noise ($S_{40} \simeq 10^{-8} $ V$^2$/Hz) pinned phase (P1) to a constant, higher-noise phase ($S_{40} \simeq 10^{-6} $ V$^2$/Hz), characteristic of the P3 phase. The dominant peak in $S_{40}$, attributed to the P2 phase at low temperatures, seen in Fig. 3(c) of the main text, is absent in Fig. S3. The zero-drive resistance in Fig. S3 is of the order of 1 M$\Omega$ and the \textit{I-V} is significantly non-linear, which can be seen from the $dV/dI$ vs. $I$ trace. The noise also increases by $\sim$ 2 orders of magnitude across the P1 to P3 transition. These observations confirm the persistence of the solid phase at this temperature and the absence of the P2 phase before the solid melts at $T$ $\simeq$ 120 mK (see section II).

\section{IV.  C\lowercase{urrent dependence of the narrow-band noise peaks}}

\begin{figure}[h]
\centering
\includegraphics[width=0.48\textwidth]{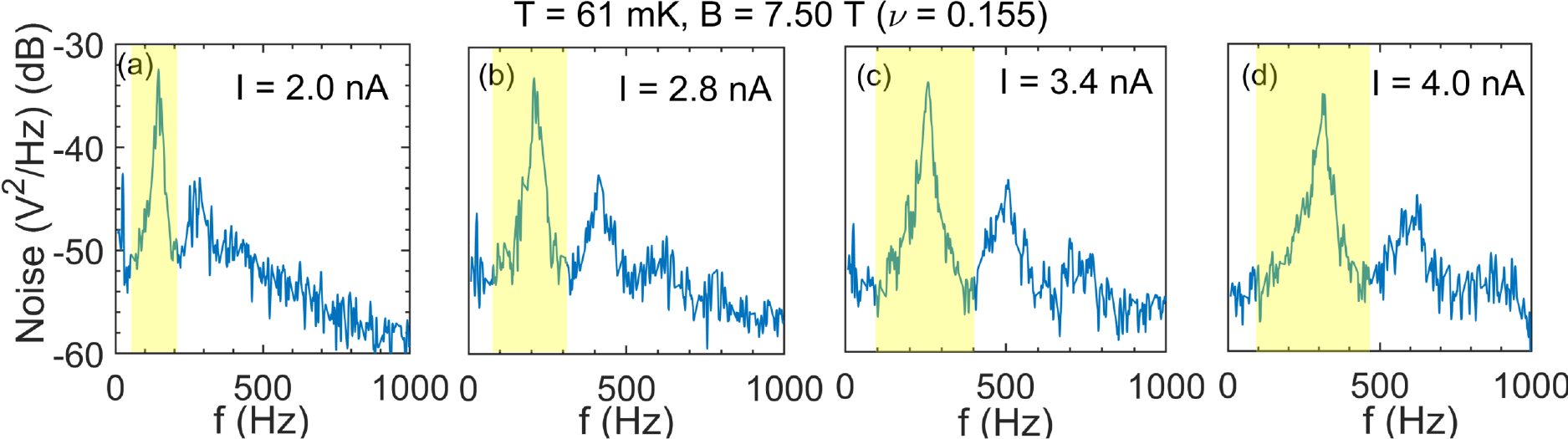}
\centering
  \caption{\label{Ip} 
 Noise power vs. $f$ for different values of $I$ as indicated, at $T$ $\simeq$ 61 mK and $B$ = 7.50 T ($\nu$ = 0.155). The yellow bands highlight the width of the first harmonic of the narrow-band noise peaks.
  }
  \label{fig:Ip}
\end{figure}

In Fig. S4, we show the noise power spectra for four different values of $I$ at $T\simeq 61$ mK, and $B$ = 7.50 T ($\nu$ = 0.155). The traces shown are deep in the P3 phase, which can be concluded from Figs. 3(a) and 3(d) of the main text. We define the width of the first harmonic of the narrow-band noise peaks as the frequency range of the yellow bands presented in Fig. S4. The widths obtained for the four different currents ($I$ = 2.0, 2.8, 3.4, and 4.0 nA) are $\simeq$ 140, 220, 300, and 370 Hz, respectively. Despite this increase in the peak width, the integrated noise power in the yellow bands stays approximately the same ($0.011 \pm 0.001$ V$^2$) for all the four current values.

\section{V. T\lowercase{emperature dependence of the narrow-band noise peaks}}

\begin{figure}[h]
\centering
\includegraphics[width=0.5\textwidth]{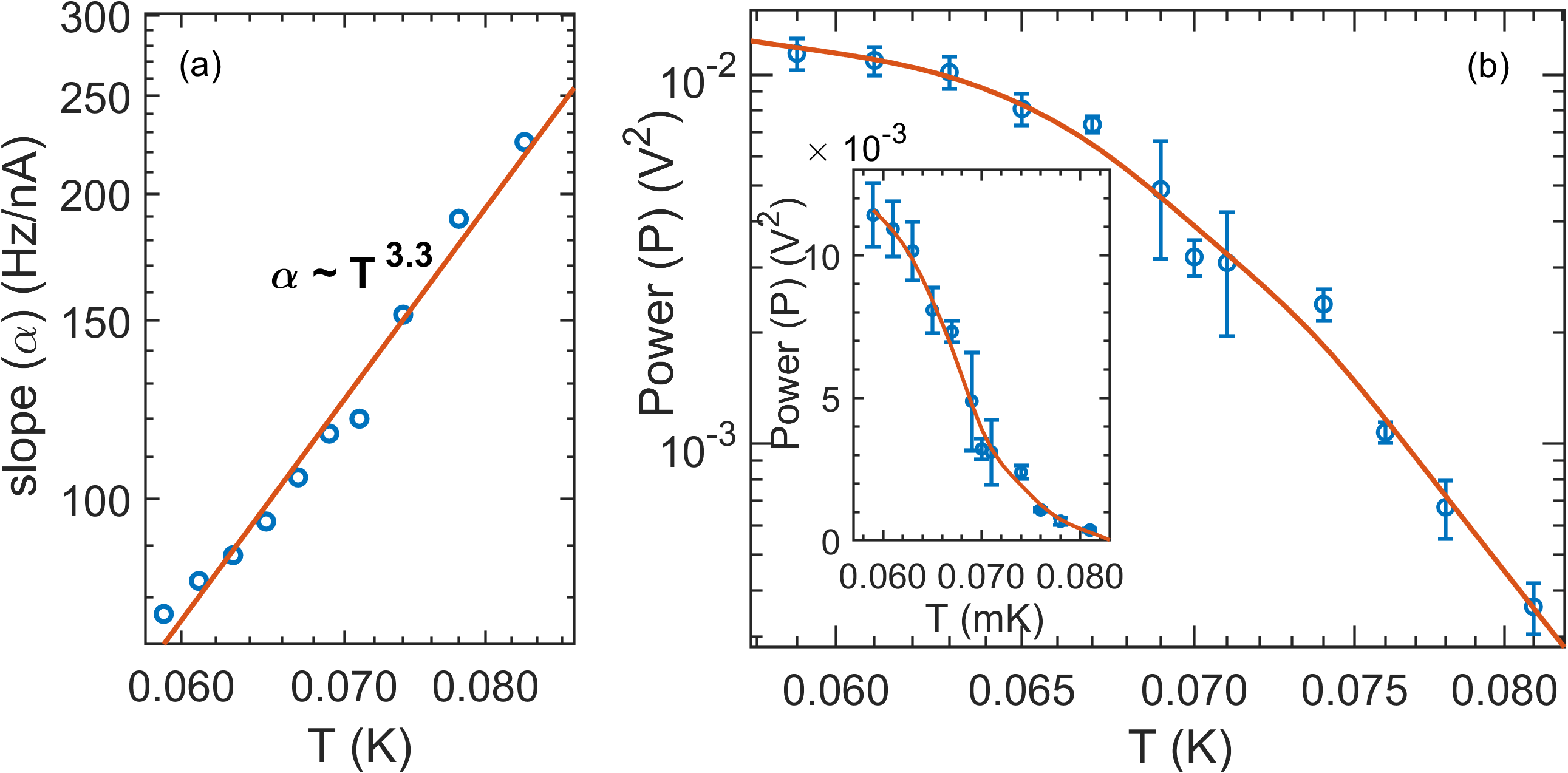}
\centering
  \caption{\label{Ip} 
 (a) Slope ($\alpha$) of the dependence of the fundamental frequency peak on $I$ in the $f$ vs. $I$ plots is shown as a function of $T$ on a log-log scale. (b) Temperature dependence of the power associated with the fundamental frequency of the narrow-band noise spectra in a log-log, and linear scale plot (inset).
  }
  \label{fig:Ip}
\end{figure}

In our measurements, we observe a strong dependence of the slope ($\alpha$) of the fundamental frequency in the $f$ vs. $I$ plots on temperature (see Figs. 3(d) and 3(e) of the main text). In Fig. S5(a), we show this dependence in a log-log plot. We observe an increasing trend of $\alpha$ with an approximate power-law dependence of $\alpha\propto T^{3.3}$. We also measure the power contained within the fundamental frequency by integrating the noise power spectrum curves in the frequency range corresponding to the yellow bands shown in Fig. S4, and then subtracting the power of the noise floor calculated for the same window. We repeat this process for different $I$ when the system is deep in the P3 phase (see, for example, Fig. S4). We then obtain the average of these power values, and the standard deviation. We present the average values in a log-log plot shown in Fig. S5(b); the error bars reflect the standard deviation. The magnitude of the slope in the log-log scale plot increases with $T$, implying that the power decreases drastically at high temperatures. We also plot the power in a linear scale, which is presented in Fig. S5(b) inset. An explanation for the temperature dependencies of $\alpha$ and the power of the first harmonic, along with the current dependence of the narrow-band noise peaks described in Section IV, awaits future theoretical work that treats the pinned WS at a microscopic level.

\section{VI.  \boldmath{$V$}  \lowercase{vs.} \boldmath{$t$} \lowercase{at} $B$ = 7.50 T ($\nu$ = 0.155) \lowercase{and} \textit{T} $\simeq$ 61 \lowercase{m}K}
\setlength{\parindent}{0mm}
\setlength{\parskip}{1mm}

\begin{figure}[h]
\centering
\includegraphics[width=0.5\textwidth]{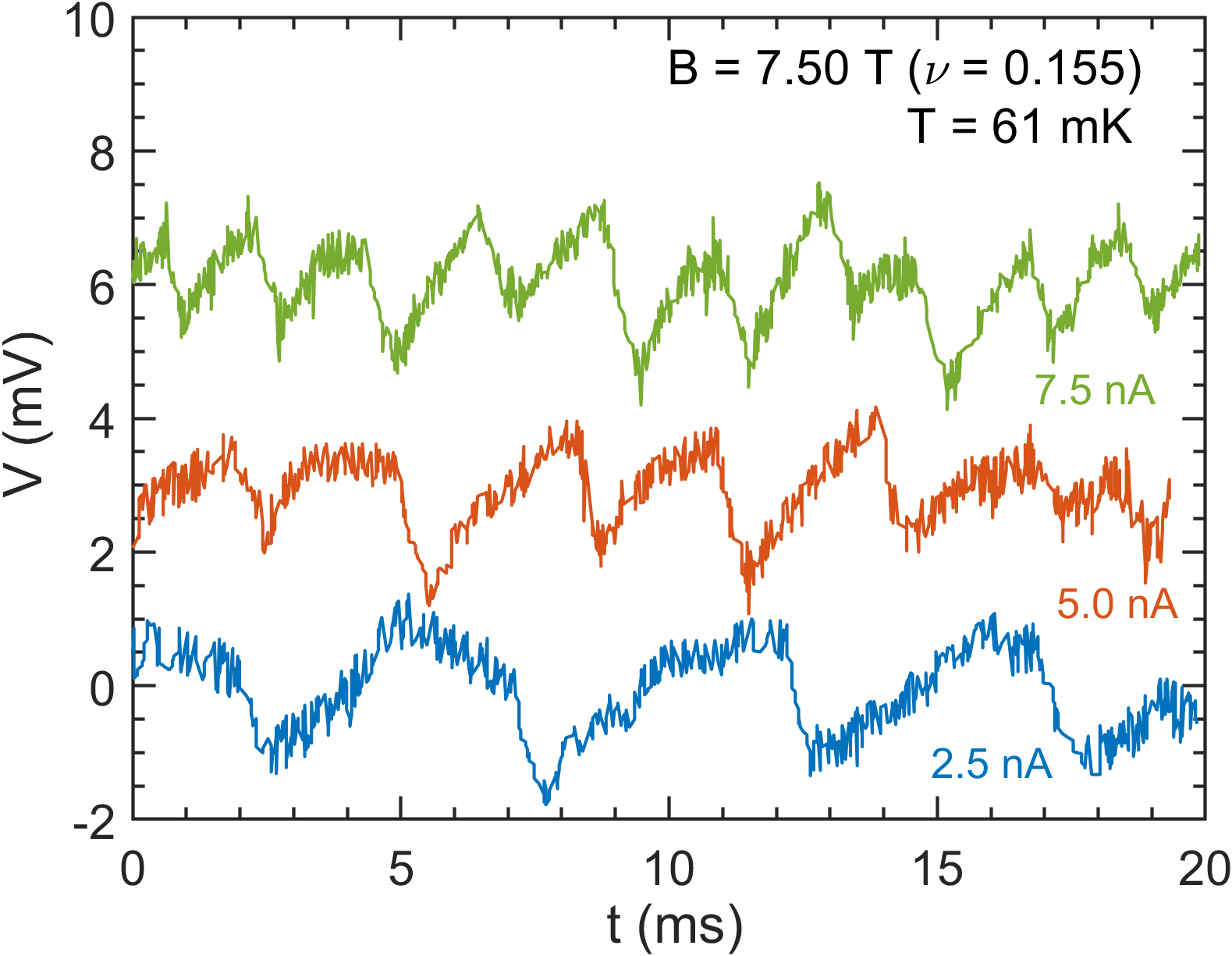}
\centering
  \caption{\label{Ip} 
 $V$ vs. $t$ for $I$ = 2.5, 5.0, and 7.5 nA at $B$ = 7.50 T ($\nu$ = 0.155) and $T$ $\simeq$ 61 mK. For clarity, he traces are shifted vertically by 0, 3, and 6 mV.
  }
  \label{fig:Ip}
\end{figure}

Figure S6 shows the $V$ vs. time ($t$) traces taken at three different currents, $I$  = 2.5, 5.0, and 7.5 nA, at 7.50 T ($\nu$ = 0.155) and $T$ $\simeq$ 61 mK. The traces, which were measured using an oscilloscope (Tektronix TDS 1002), are shifted vertically by 0, 3, and 6 mV, respectively. The signal was $ac$-coupled at the pre-amplifier stage and so, the $dc$ voltage was filtered out, as shown in Fig. S1(c). For $I$ mentioned above, the WS is deep in the P3 phase, which can be seen from Fig. 3(a) of the main text. 

The waveform of the real-time signals is saw-tooth like, comprised of a slow rise and a fast drop, suggesting a stick-slip motion of the WS, similar to what has been previously reported for the sliding dynamics of the charge density waves (CDWs) \cite{gruner1994density,monceau2012electronic}. However, because of an uncertainty in determining the polarity of the $V(t)$ traces in the experimental setup, it is possible that the saw-tooth profile is flipped, in which case there would be a fast rise followed by a slow decay. This type of waveform has also been previously observed in CDW motion and is attributed to the stick-slip motion of the CDW as it slides over a disordered background \cite{sekine2004sliding}. 

\section{VII. R\lowercase{eproducibility of the non-linear} \textit{I-V} \lowercase{data}}
\begin{figure}[h]
\centering
\includegraphics[width=0.475\textwidth]{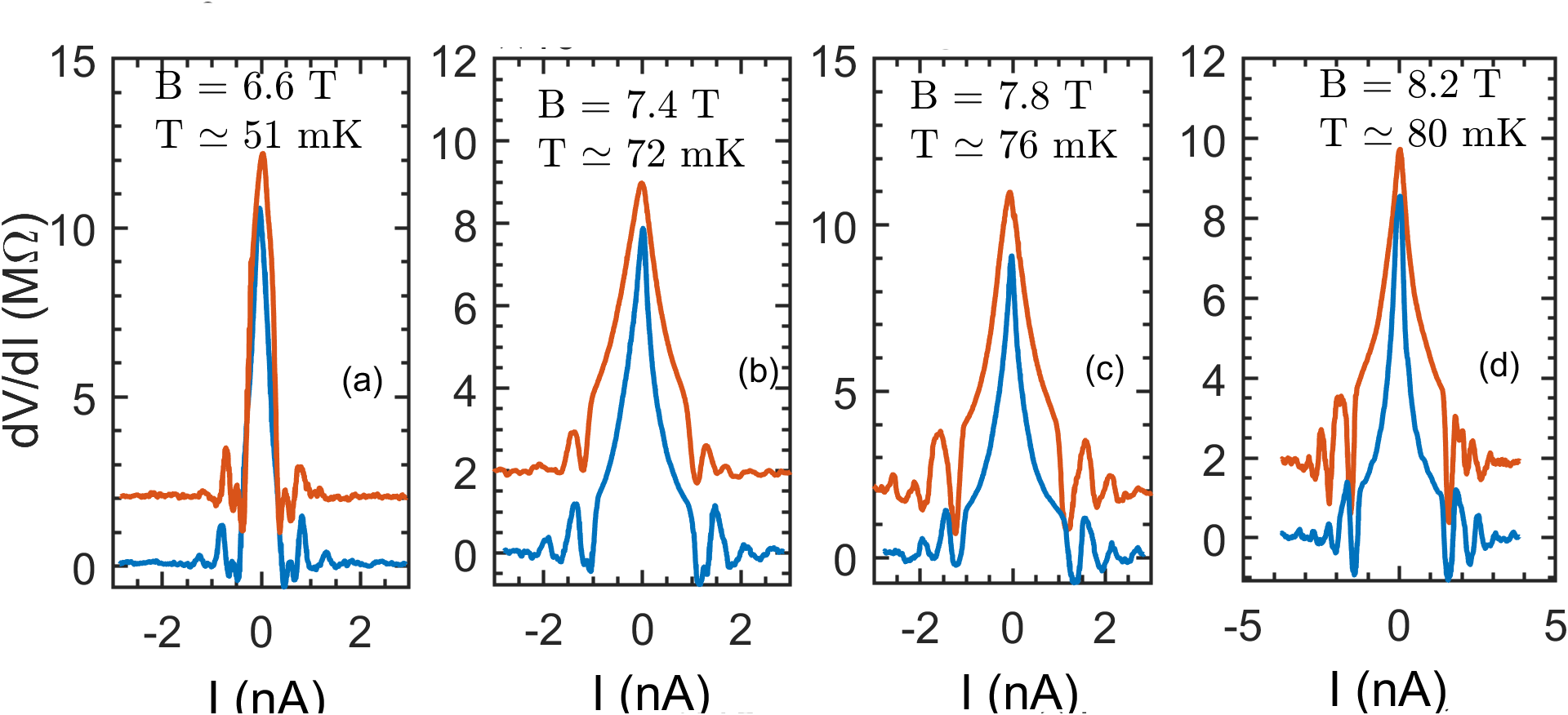}
\centering
  \caption{\label{Ip} 
Reproducibility of the non-linear \textit{I-V} data. In each panel, the blue trace was taken after a thermal cycling to 6 K. The red traces are shifted up by 2 M$\Omega$. The temperatures at which the two traces in each panel were taken are very close but not identical (see text).
  }
  \label{fig:Ip}
\end{figure}

In this section, we discuss the reproducibility of the data after thermal cycling. In each panel of Fig. S7 we show two traces (blue and red) taken at a given magnetic field. Each blue trace is obtained after warming the sample up to $T=6$ K and then cooling it back to the same low temperature as in the red trace, as indicted in each panel. The red traces are shifted up by 2 M$\Omega$ for clarity and ease of comparison. We note that each pair of traces, although not identical, follow each other closely. They show similar threshold currents, and the subsequent fluctuations are also repeatable. There are however, some quantitative differences, which can be attributed to the fact that the new state of the sample after a thermal cycle is not identical to the old state. Also, the temperatures at which the two traces were measured in each panel are not exactly the same (e.g. the blue and the red traces at $B = 8.2$ T in panel (d) were taken at nominal temperatures of 79.6 mK and 80.6 mK, respectively).